\documentstyle[aps,prb,multicol,floats,psfig]{revtex}

\begin{document}
\newcommand{\be}{\begin{equation}}
\newcommand{\ee}{\end{equation}}
\draft
\preprint
\widetext
\title {\bf
Metal-to-insulator transition and magnetic
ordering
 in CaRu$_{1-x}$Cu$_x$O$_3$ }

\author
 {I. M. Bradari\'c,
$^{a,}$
\footnote{ Corresponding author. Fax: +381 11 3440 100,
 e-mail: bradaric@rt270.vin.bg.ac.yu}
I. Felner$^b$ and M. Gospodinov$^c$}
\address{$^a$ Laboratory for Theoretical and
Condensed Matter Physics, The "Vin\v ca" Institute of Nuclear Sciences\\
 P.O. Box 522, 11001 Belgrade, Serbia,
Yugoslavia.\\
 $^b$ The Racah Institute of Physics, The Hebrew
University of Jerusalem,\\
 Jerusalem 91904, Israel.\\
 $^c$ Bulgarian Academy of Sciences, Institute of
Solid State Physics,\\
 72 Tzarigradsko Chaussee Boulevard, Sofia 1784, Bulgaria.}
\date{\today}
\maketitle
\widetext
\begin{abstract}

\vskip 3mm

CaRuO$_3$ is perovskite with an orthorhombic distortion and is
believed to be close to magnetic ordering. Magnetic studies of
single crystal and polycrystalline CaRu$_{1-x}$Cu$_x$O$_3$ ($0\le x
\le 15$~at.\%~Cu) reveal that spin-glass-like transition
develops for $x\le 7$~at.\%~Cu and obtained value for
effective magnetic moment $p_{\rm eff}=3.55\,\mu_B$ for $x=5$~at.\%~
Cu, single crystal, indicates presence of Ru$^{5+}$. At higher Cu concentrations
more complex magnetic behaviors are observed. Electrical
resistivity measured on polycrystalline samples shows
metal-to-insulator transition (MIT) at 51 K for only 2~at.\%~
Cu. Charge compensation, which is assumed to be present upon
Cu$^{2+/3+}$ substitution, induces appearance of Ru$^{5+}$
and/or creation of oxygen vacancies in crystal structure. Since
the observed changes in physical properties are completely
attributable to the charge compensation, they {\it cannot be related}
to behaviors of pure compound where no such mechanism is
present. This study provides the criterion for ``good" chemical
probes for studying Ru-based perovskites.

\end{abstract}


\vskip 0.8cm

\section{INTRODUCTION}

\vskip 1.2 true cm

Following the discovery of superconductivity in Sr$_2$RuO$_4$ at
$\approx 1$ K$^1$ the wide interest in Ru-based perovskites
has been generated, due to diversity of unusual physical
properties discovered. Structurally related Ca$_2$RuO$_4$, which
has the same crystal structure as high~$T_c$ superconductor
La$_{2-x}$Sr$_x$CuO$_4$, is nonmetallic and shows antiferromagnetic
(AFM) ground state below $T_N=110$ K.$^{2-4}$ On the other
hand SrRuO$_3$ and CaRuO$_3$ ($n=\infty$ members of the alkaline
earth-ruthenium Ruddlesden-Popper$^5$ series
(Sr,Ca)$_{n+1}$Ru$_n$O$_{3n+1}$) are exceptionally interesting {\it
per se}. Both compounds adopt the same perovskite structure with
an orthorhombic distortion (GdFeO$_3$ structure type) and are
metallic conductors. SrRuO$_3$ is the only known ferromagnetic
(FM) conductor among the $4d$ oxides ( Curie temperature
$T_c=165\rm\,K$ ), whereas CaRuO$_3$ has been recently shown to have
spin-glass-like magnetic ground state.$^6$ Since a common
structural feature of the two compounds is that they are composed
of an array of corner-shared octahedra RuO$_6$, it is assumed
that the degree of tilting and rotation of these octahedra from
ideal cubic-perovskite structure governs the observed
differences in the magnetic ground states. A narrow itinerant
$4d$ band is formed through hybridization of Ru~$t_{2g}$ and
O~$2p$ orbitals. The $4d$ bandwidth thus formed sensitively
depends on degree of hybridization.$^7$ One of powerful tools to
study physical properties of such systems is realized through
chemical substitution. Recent results on the effects of chemical
substitutions in CaRuO$_3$ show following:
Ca$_{0.95}$Na$_{0.05}$RuO$_3$$^8$ spin-glass or AFM ordering at 55
K, CaRu$_{1-x}$Sn$_x$O$_3$$^9$ spin-glass ordering for $4\le x\le
10$~at.\%~Sn and MIT for $x\ge 16$~at.\%~Sn and
CaRu$_{1-x}$Rh$_x$O$_3$$^{10}$ magnetic ordering (spin-glass?) for
all $x$ and MIT for $x\ge 7$~at.\%~Rh. From these studies it
is evident that physical properties of CaRuO$_3$ are much more
influenced by Rh than Sn substitution.  Substitution of
nonmagnetic Sn$^{4+}$ for Ru$^{4+}$ represents only lattice
frustration and magnetic dilution of Ru-sublattice. On the
other hand, Rh$^{4+}$($4d^5\!\!: t_{2g}^5 e_g^0$) with $S=1/2$ in the
low-spin state, behaves as magnetic impurity, since it substitutes
Ru$^{4+}$($4d^4\!\!: t_{2g}^4 e_g^0$) with low-spin $S=1$.
Nevertheless, it is likely that Rh assumes valence state
Rh$^{3+}$($4d^6\!\!: t_{2g}^6 e_g^0$), $S=0$, when incorporated in
Ru-sublattice and thus produces charge frustration of the
system. In order to clarify this issue, we have chosen
Cu$^{2+}$($3d^9$), $S=1/2$, as substitution for Ru$^{4+}$, since
this inevitably represents both charge and spin frustration.
CaCuO$_2$ is considered as the parent of the cuprate family of
superconducting compounds, consisting of CuO$_2$ sheets with AFM
ordering of Cu$^{2+}$ neighboring cations. Although CaRuO$_3$
and CaCuO$_2$ are not isostructural, we have assumed that
appreciable amount of Cu can be incorporated in
Ru-sublattice, while preserving the crystal structure.
Furthermore, recent discovery of coexistence of magnetism and
superconductivity in ruthenium-based layered cuprates
$Ln_{1.4}$Ce$_{0.6}$RuSr$_2$Cu$_2$O$_{10-\delta}$ and $Ln$RuSr$_2$Cu$_2$O$_8$
($Ln$=Eu, Gd)$^{11-14}$ proves an added motivation for
this study.

In this paper we report results of magnetic and electrical
resistivity properties of CaRu$_{1-x}$Cu$_x$O$_3$ ($0\le x\le 15$~at.\%~Cu).
We show here that Cu substitution on Ru sites
profoundly alters ground state properties of CaRuO$_3$ introducing
mixed oxidation states of Ru cations.

\vskip 1.2 true cm

\section{EXPERIMENTAL DETAILS}

\nobreak

\vskip 1.2 true cm

Polycrystalline samples of CaRu$_{1-x}$Cu$_x$O$_3$ ($0\le x\le 15$~
at.\%~Cu) were prepared by solid-state reaction from the
appropriate stoichiometric mixtures of Ru metal powder, CaCO$_3$ and CuO
(purchased from Strem Chemicals Inc.). The samples were mixed, ground
and preheated at $850^\circ\rm\,C$ for 24 h in air. The powders
were then  reground,
pressed into pellets and heated at $1000-1200^\circ\rm\,C$ for 72 h in air,
with two intermediate grindings. Single crystals of CaRu$_{1-x}$Cu$_x$O$_3$
($x=0$, 5, 10 and 15 at.\%~Cu) were grown from ground mixture of sintered
samples and CaCl$_2$ flux in the ratio 1:30 (sample : flux). The mixture
was heated to $1260^\circ\rm\,C$ and soaked at this temperature
for 24 h, followed
by slow cooling at a rate $2^\circ\rm\,C/h$ to $1000^\circ\rm\,C$ and
finally  quenched to
room temperature to avoid possible twinning. The crystals tend to form in
almost square planar shape with sizes around
$0.4\times0.4\times0.02\rm\,mm^3$,
with short dimension along $c$-axis. The single crystals and sintered
polycrystalline samples were characterized by energy dispersive x-ray
analysis (EDAX),
scanning electron microscopy (SEM) and powder x-ray diffraction (XRD) measurements.
Powder x-ray diffraction measurements were performed on a Philips 1010 powder
diffractometer using Cu~K$\alpha$ radiation at room temperature. DC magnetic measurements were
performed  by a Quantum Design superconducting quantum interference
device (SQUID) magnetometer. Resistivity measurements were performed on
polycrystalline samples employing standard four-point method. Unfortunately, due to extreme
fragility of single crystals, several attempts to measure resistivity were not successful.


\section{RESULTS AND DISCUSSION}

\nobreak


The samples are of single phase and crystallize in perovskite
structural type with orthorhombic structure, space group {\it Pnma}
(62).  Our results for the lattice parameters of single crystal
and polycrystalline samples are in excellent agreement with
values previously published.$^{15, 16}$ At concentrations greater
than 15~at.\%~Cu, small impurity diffraction lines of CuO
appear so that it is assumed that at this concentration the
solubility limit is reached. The concentration dependence of
the room temperature lattice parameters and the volume of the
unit cell for polycrystalline samples are shown in Fig. 1.
The data for $a$ and $b$ axes show the expected Vegard's law
linear  expansion assuming that larger Cu$^{2+}$
(ionic radius 0.73 \AA) substitutes Ru$^{4+}$
(ionic radius 0.62 \AA). On the other hand, $c$-axis
is almost constant below 2~at.\%~Cu and then shows a linear
shrinkage with increasing Cu concentration. Similar behavior
has been observed in La$_{2-x}$Sr$_x$Cu$_{1-y}$Ru$_y$O$_{4-\delta}$$^{17,
18}$ for $x=2$ and $0.7\le y\le 1.0$\,. Nevertheless, the volume
of the unit cell inreases with $x$, as expected.

Shown in Fig. 2 is the magnetization vs temperature
for CaRu$_{0.95}$Cu$_{0.05}$O$_3$ single crystal measured
at various values of magnetic field applied along [001] direction,
under field-cooled (FC) and zero-field-cooled (ZFC)
conditions. An irreversibility phenomenon associated
with spin-glass-like behavior in CaRuO$_3$ (see inset of Fig. 2) is clearly
present in this case. Note the strong field dependence
of the irreversibility temperature $T_{\rm irr}$ (defined as the
divergent point in the ZFC and FC curves). Furthermore,
the broad peak around 35 K in the ZFC curve at 100 Oe is
shifted to 29 K at 500 Oe and completely smeared at 1 kOe.

The observed spin-glass behavior can be accounted for by
assuming the following simple model: Cu$^{2+}$ and/or Cu$^{3+}$
substitution for Ru$^{4+}$ produces charge frustration of
the system and therefore requires partial oxidation of neighboring
Ru$^{4+}$ cations to higher valence state Ru$^{5+}$ ($4d^3$).
This frustration is not local since the next-to-nearest-neighbors
will also be affected, undergoing partial reduction to Ru$^{3+}$($4d^5$).
In this manner, charge compensation mechanism may affect the
valence state and therefore the magnetic moment of Ru cations at
appreciable distance from Cu cation, simultaneously reducing the
number of available conduction paths. According to Ref. 19,
the signs of transfer integrals for $180^\circ$ cation-anion-cation
(CAC) superexchange interactions between octahedral-site cations
are predicted to be ferromagnetic for all Cu$^{2+/3+}$-Ru$^{3+/4+/5+}$
combinations and antiferromagnetic for Cu$^{2+}$-Cu$^{2+}$,
Ru$^{5+}$-Ru$^{3+}$(weak), Ru$^{4+}$-Ru$^{3+}$(weak) and
Ru$^{5+}$-Ru$^{5+}$ combinations. Therefore, mixed interactions
together with random distribution of Cu atoms within Ru-sublattice
create necessary ingredients for spin-glass ordering.
Within the framework of the proposed scenario, ferromagnetic clusters
are formed around Cu impurities with AFM interactions within and
between clusters, introducing frustration that result in a
spin-glass behavior.

We fitted the data of $\chi(T)=M(T)/H$ (single crystal) in the paramagnetic range
($120\le T\le 250 \rm\,K$) to the modified Curie-Weiss (CW) law
$\chi(T)=\chi_0+C/(T-\theta)$, where $\chi_0$ is the temperature
independent term, $C$ is the Curie constant and $\theta$ is the
CW temperature. The value of the effective magnetic moment
(deduced from the Curie constant $C$) $p_{\rm eff}=3.55\,\mu_B$ is
remarkably close to the expected Hund's rule value $3.87\,\mu_B$
for Ru$^{5+}$ with $S=3/2$ and $g=2$, indicating presence of Ru$^{5+}$ in this
mixed-valent system.
This is  in marked contrast to $p_{\rm eff}=2.33\,\mu_B$ obtained
for CaRuO$_3$ single crystal,$^6$ appropriate to $2.83\,\mu_B$
expected for low-spin state ($S=1$) Ru$^{4+}$. The Curie-Weiss
temperature $\theta$ also drastically changes from $-36(1)\rm\,K$ for
CaRuO$_3$ to $-134(3)\rm\,K$ for CaRu$_{0.95}$Cu$_{0.05}$O$_3$, showing
enhanced AFM interactions. Furthermore, $\chi_0$ a measure of the
density of states near the Fermi surface drops from $\chi_0=9.5\times10^{-3}
\rm\,emu/mole\,Oe$ at $x=0$ to $\chi_0=7\times10^{-4}\rm\,emu/mole\,Oe$
at $x=0.05$. This is consistent with resistivity behavior measured
on polycrystalline samples shown in Fig. 7, where the onset of
MIT is observed at 51 K and  69 K for
$x=0.02$ and $x=0.05$ Cu concentrations, respectively.

Although these results support the charge compensation mechanism
outlined above, yet another possibility for charge compensation
realized by oxygen loss, i.e. creation of oxygen vacancies,
can be significant. It has been shown for
La$_{2-x}$Sr$_x$Cu$_{1-y}$Ru$_y$O$_{4-\delta}$,$^{18}$ with K$_2$NiF$_4$-type
structure, that the oxygen vacancies are located exclusively
in the vicinity of Cu cations, which consequently means local
character of charge compensation. However, we believe that oxygen
loss is not significant in our samples for $x\le7$~at.\%~Cu,
since the lower concentration of Ru$^{5+}$, due to presence of oxygen
vacancies, would increase conductivity and shift MIT to higher
Cu concentrations, simultaneously reducing the MIT temperature for fixed $x$.
Indeed, resistivity measurements (not shown here) performed on
slightly reduced polycrystalline samples (annealed at $350^\circ\rm\,C$
in N$_2$ atmosphere for an hour) show metallic behavior in the
measured temperature range $7\le T\le 300\rm\,K$ for $x\le 3$~at.\%~Cu
and decrease of  MIT temperature  from 87 K to 30 K for
$x=7$ at.\% Cu.

Magnetic isotherms $M(H)$ at $T=5\rm\,K$ obtained after cooling
in zero applied field for an almost square planar shaped
CaRu$_{0.95}$Cu$_{0.05}$O$_3$ single crystal and polycrystalline
samples are shown in Fig. 3. With increasing applied field along $c$-axis,
$M(H)$ reaches saturation for $H\ge2\rm\,T$,
yielding very low saturation moment $p_0=0.044\,\mu_B$.
On the other hand, $M(H)$ for field applied perpendicular to $c$-axis
(i.e. in plane), shows no sign of saturation up to $H=5\rm\,T$.
This behavior is markedly different from the anisotropy of the
magnetization found in single crystal CaRuO$_3$,$^6$ where the
easy axis of magnetization along $c$-axis is determined.
Isothermal magnetization obtained for polycrystalline
CaRu$_{0.95}$Cu$_{0.05}$O$_3$ sample shows linear behavior
with no qualitative difference from that obtained for
polycrystalline CaRuO$_3$. Small hysteresis loops at $T=5\rm\,K$
(not shown here) with essentially same coercive field of
$H_c\approx100\rm\,Oe$
for both polycrystalline and single crystal samples are observed.

The temperature dependence of the magnetic susceptibility for
polycrystalline (a) and $M(T)$ for single crystal
(b) CaRu$_{0.9}$Cu$_{0.1}$O$_3$ are shown in Fig. 4.
Apart from different irreversibility temperatures $T_{\rm irr}$,
observed for polycrystalline and single crystal materials
(also present for $x=0$ and 5 at.\% Cu), truly remarkable difference
is manifested here both in shape of FC and ZFC curves and the onset
of magnetic ordering  at $T_0$ for polycrystalline sample,
but without any magnetic anomaly present in single crystal
in this temperature range. $\chi(T)$ data for polycrystalline
CaRu$_{0.9}$Cu$_{0.1}$O$_3$ sample show $T_{\rm irr}\approx 80\rm\,K$
and well defined maximum in ZFC curve at $T_0\approx 9.2\rm\,K$
($H=14\rm\,Oe$), Fig. 4a. At $H=1\rm\,kOe$, $T_{\rm irr}$
is completely suppressed being essentially equal to $T_0$,
which is slightly shifted to lower temperatures at this field.
On the other hand, $M(T)$ data for single crystal
CaRu$_{0.9}$Cu$_{0.1}$O$_3$ resemble re-entrant spin glass
(RSG) behavior. Note the pronounced FM-like shape of FC curves
and also difference between $T_{\rm irr}$ and the onset of FM-like behavior.
$T_{\rm irr}$ is strongly field dependent, decreasing rapidly with
increasing field.

The increase of Cu concentration promotes the growth of FM clusters
tending to establish long-range FM order. However, this process is
opposed by two effects: 1) The influence of short range AFM interactions
within and between FM clusters may become more prominent in sintered
polycrystalline sample with randomly oriented microcrystallites than
in macroscopic single crystal, thereby leading to freezing of magnetic
moments at low temperatures. Spin disorder at grain boundaries and surfaces
has been extensively studied recently,$^{20-23}$ including the size-dependent
magnetic properties.$^{24}$ We believe that these effects are present in this
case, specially because the material shows intrinsic spin-glass properties.
2) The decrease of the average
distance between Cu cations with increasing concentration leads to
increased concentration of Ru$^{5+}$ cations and thus higher probability
for  Ru$^{5+}$-Ru$^{5+}$ AFM superexchange interactions. Formation
of Cu-O-Cu pairs, triplets etc. and creation of oxygen vacancies at
higher Cu concentrations add to complexity of magnetic properties.
The former effect provides possible explanation for differences in
magnetic properties between polycrystalline and single crystal
CaRu$_{0.9}$Cu$_{0.1}$O$_3$. The growth of FM clusters is almost
unrestricted in single crystal, which is reflected in extremely
large coercive field $H_c=2.2\rm\,T$ (see Fig. 5a).
Saturation moment $p_0=0.14\,\mu_B$ deduced from $M(H)$ curve
is significantly higher than $p_0=0.044\,\mu_B$
obtained for CaRu$_{0.95}$Cu$_{0.05}$O$_3$. The extracted
paramagnetic values in the range $120\le T\le 250\rm\,K$ are:
$\chi_0=3.6\times10^{-4}\rm\,emu/mole\,Oe$, $\theta=-219\rm\,K$ and
$p_{\rm eff}=3.48\,\mu_B$.

The increase of Cu concentration from 5 to 10~at.\%~Cu
also leads to profound changes in $\rho(T)$ behavior (see Fig. 7),
showing sharp increase of the MIT temperature from 87 K to above 300 K
for $x=7$ and 10 at.\%~Cu, respectively.

Shown in Fig. 6 are the magnetic susceptibility $\chi(T)$ vs
temperature dependence for polycrystalline (a) and $M(T)$ for
single crystal (b), CaRu$_{0.85}$Cu$_{0.15}$O$_3$ samples.
Both effects mentioned above should be considered in this case,
since the low temperature magnetic anomaly at $T_0$,
associated with freezing of FM clusters, is present in polycrystalline
and single crystal material. However, note the change of shape of FC
curve for polycrystalline sample at $H=1\rm\,kOe$ and also change
of slope at $T_0$, which are not seen in single crystal.
The onset of FM-like behavior at 70-80 K (FC curves)
is very close for both forms of material and is
virtually field independent for single crystal.
Moreover, $T_{\rm irr}$ is also approximately equal
and decreases rapidly with increasing field in
similar manner. The behavior of low temperature magnetic
anomaly observed in ZFC curves vs magnetic field is very
similar for both single crystal and polycrystalline samples
and shows slight temperature decrease with increasing field, being completely
washed out at $H>7.5\rm\,kOe$ in single crystal.

Magnetic hysteresis loop measured on single crystal after ZFC
(see Fig. 5b) is drastically smaller, $H_c=40\rm\,Oe$,
than that seen for $x=10$~at.\%~Cu, while saturation moment
$p_0=0.2\,\mu_B$ is somewhat larger, in accordance with differences
in magnetic properties. Fitting of $\chi(T)$ to the CW law within
$120\le T\le 250\rm\,K$ range yields: $\theta=-33.9\rm\,K$,
$p_{\rm eff}=2.24\,\mu_B$ and $\chi_0=4.8\times10^{-3}\rm\,emu/mole\,Oe$.
The higher Cu concentration, inevitable presence of oxygen vacancies
at these concentrations and the temperature range of fitting
(vicinity of the magnetic transition at $\approx 80\rm\,K$)
may account for the striking change of the extracted paramagnetic
values compared with those obtained for $x=5$ and 10~at.\%~Cu
single crystals.

Shown in Fig. 7 are the electrical resistivity data vs temperature
for polycrystalline CaRu$_{1-x}$Cu$_x$O$_3$ samples. Although there is
a close temperature correspondence between changes of slope of 
$d(\ln\rho/d(1/T^4)$ vs. $1/T^4$ and $T_{\rm irr}$ ($T_0$) for
low applied magnetic field M(T) data (polycrystalline samples), the plots
for $\ln\rho$ vs. temperature to the $-\nu$ th power $(1\le \nu \le 1/4)$ 
do not show linear parts in any temperature regions for any $\nu$ values.
Therefore, formulae for variable-range-hopping (VRH) resistivity do not match
exactly with the resistivity data. This is not surprising, considering the large differences 
in magnetic properties between single crystal and polycrystalline samples due to 
grain boundaries.

More detailed research, employing precise control of oxygen
stoichiometry and complementary experimental methods, would
provide the basis for full characterization of the CaRu$_{1-x}$Cu$_x$O$_3$
system. However, the main result of this study shows that substitution
of Cu$^{2+/3+}$ cation for Ru$^{4+}$ changes the valence state of
the latter to Ru$^{5+}$, thereby introducing drastic changes
in both resistivity and magnetic properties of the parent compound.
Keeping in mind the obvious differences, similarity in
$\rho(T)$ and $\chi(T)$ behavior with CaRu$_{1-x}$Rh$_x$O$_3$$^{10}$
is quite appealing. While the spin-glass-like behavior in
CaRu$_{1-x}$Cu$_x$O$_3$ (for $x>0$) can be understood,
at least qualitatively, the origin of the spin-glass-like transition
observed in pure CaRuO$_3$ (lacking the evident source of
perturbation) must be more subtle in nature and is {\it not related}
to the former. Therefore, we conclude that chemical probes,
which potentially alter the valence state of Ru$^{4+}$ in
(Ca,Sr)RuO$_3$, are not adequate tools for obtaining deeper
insight in physical properties of the parent compound.
In this sense nonmagnetic probe like Sn$^{4+}$ can provide
much more information. Discussion of this issue is beyond
the scope of this article and will be published elsewhere.

\vskip 1.2 true cm

\section{CONCLUSIONS}

\nobreak

In summary, we have shown that substitution of Cu$^{2+/3+}$
for Ru$^{4+}$ in CaRu$_{1-x}$Cu$_x$O$_3$ alters the oxidation
state of the neighboring Ru cations to Ru$^{5+}$, leading
to spin-glass-like behavior for lower Cu concentrations
($x\le 7$~at.\%~Cu) and complex magnetic behaviors for
$x\ge 10$~at.\%~Cu. Simultaneously, MIT is observed for only
2~at.\%~Cu at 51 K. Physical processes involved, while being
interesting {\it per se} and possibly useful for understanding
complexities of other materials like GdRuSr$_2$Cu$_2$O$_8$,$^{14}$
have different origin and therefore cannot be related to
physical properties of the parent compound CaRuO$_3$.


\acknowledgments
We are indebted to Dr. D. Rodi\' c for help in XRD analysis.
I. M. Bradari\'c gratefully acknowledges support from the "Abdus Salam" ICTP,
Trieste, Italy.
The work in Jerusalem was supported by the BSF (1999).

\newpage

\vskip 1.2 true cm

\begin{figure}[h]
\centerline{\psfig{file=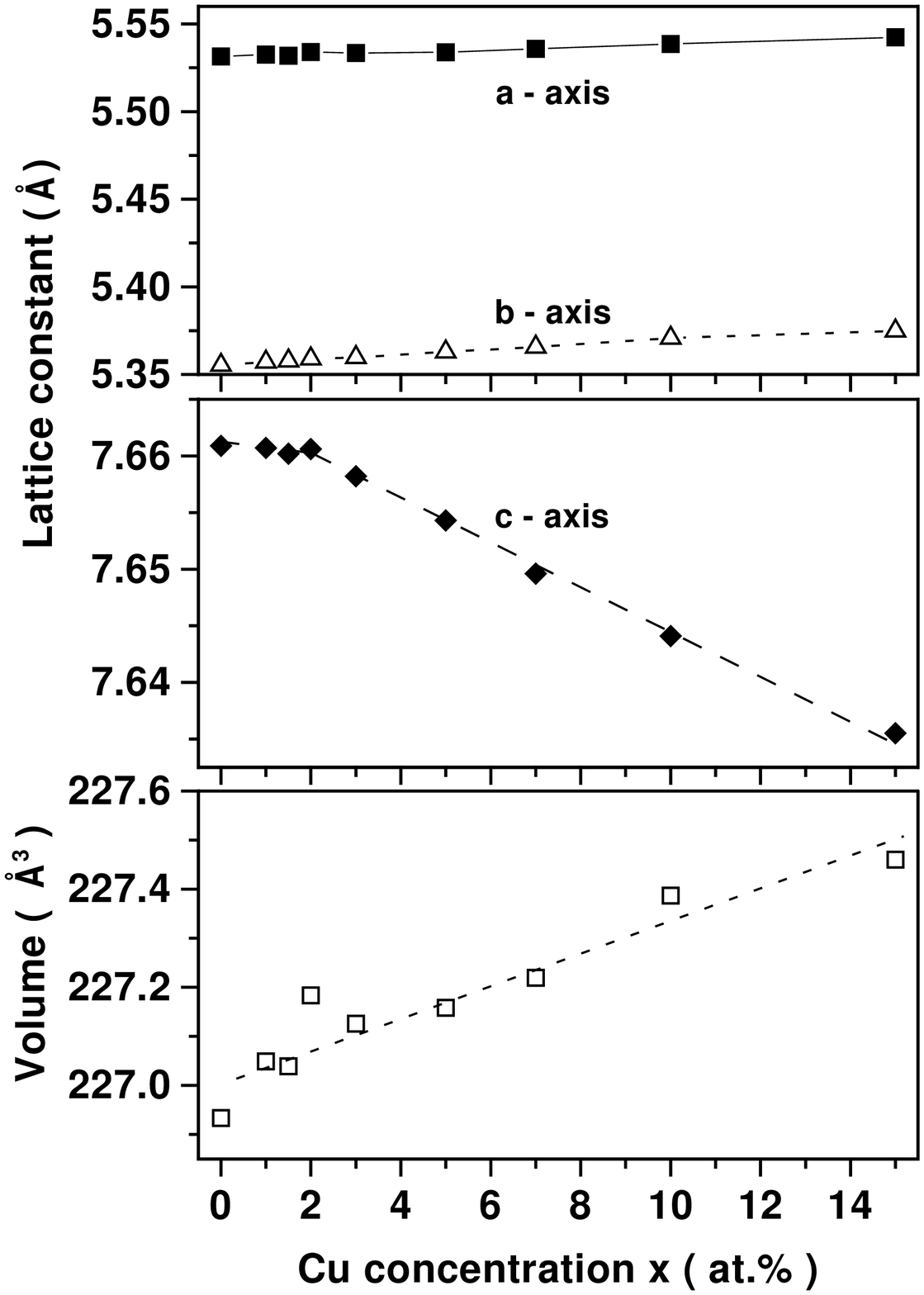,width=15cm}}

\caption{
Room temperature lattice parameters and the volume of the
unit cell vs Cu concentration for polycrystalline CaRu$_{1-x}$Cu$_x$O$_3$.}
\label{fig1}
\end{figure}

\begin{figure}[h]
\centerline{\psfig{file=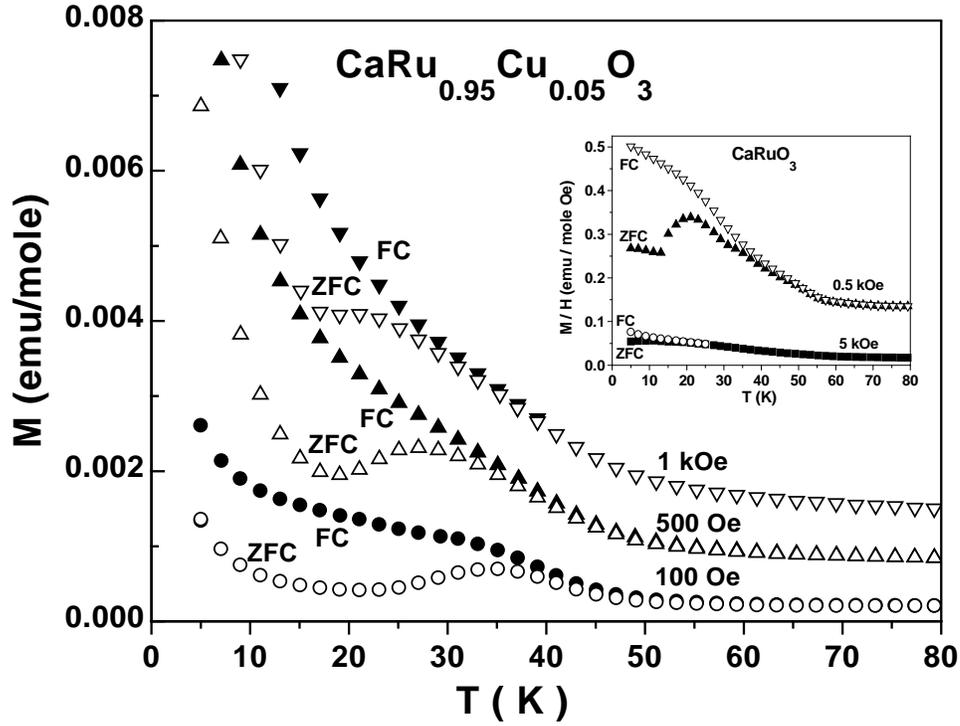,width=15cm}}

\caption
{$M(T)$ vs $T$ (FC and ZFC) curves for
CaRu$_{0.95}$Cu$_{0.05}$O$_3$ single crystal measured
at different magnetic fields applied along $c$-axis.
Inset: $\chi(T)=M(T)/H$ vs $T$ for CaRuO$_3$ single crystal
($H \bot$ $c$-axis).$^6$}
\label{fig2}
\end{figure}

\begin{figure}[h]
\centerline{\psfig{file=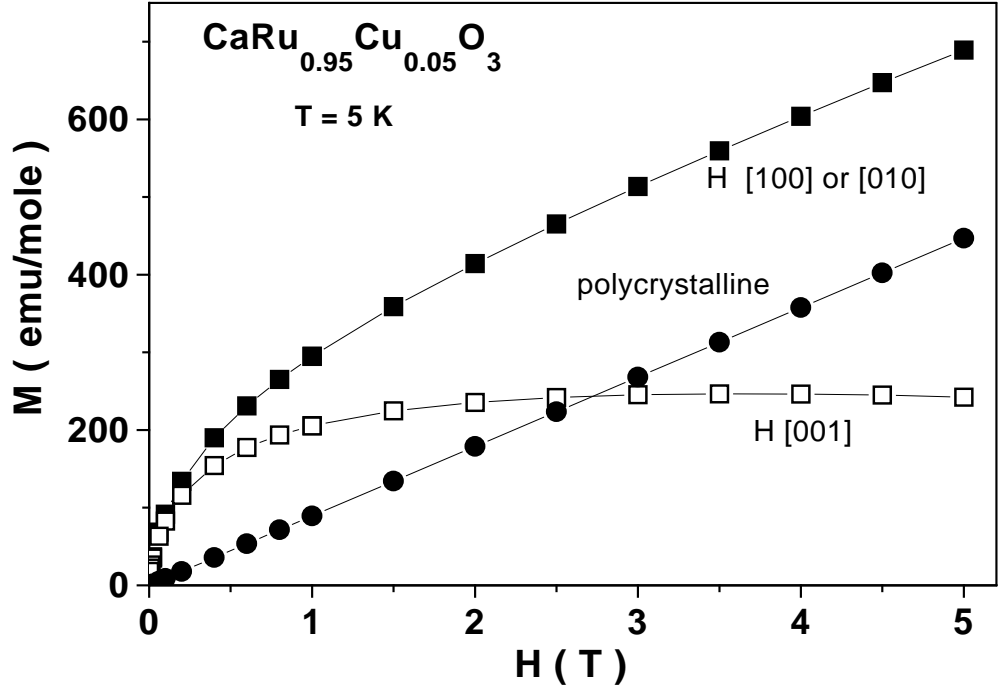,width=15cm}}

\caption
 {Magnetic isotherms at $T=5\rm\,K$ for $x=5$ at.\%~Cu
single crystal and polycrystalline materials.}
\label{fig3}
\end{figure}

\begin{figure}[h]
\centerline{\psfig{file=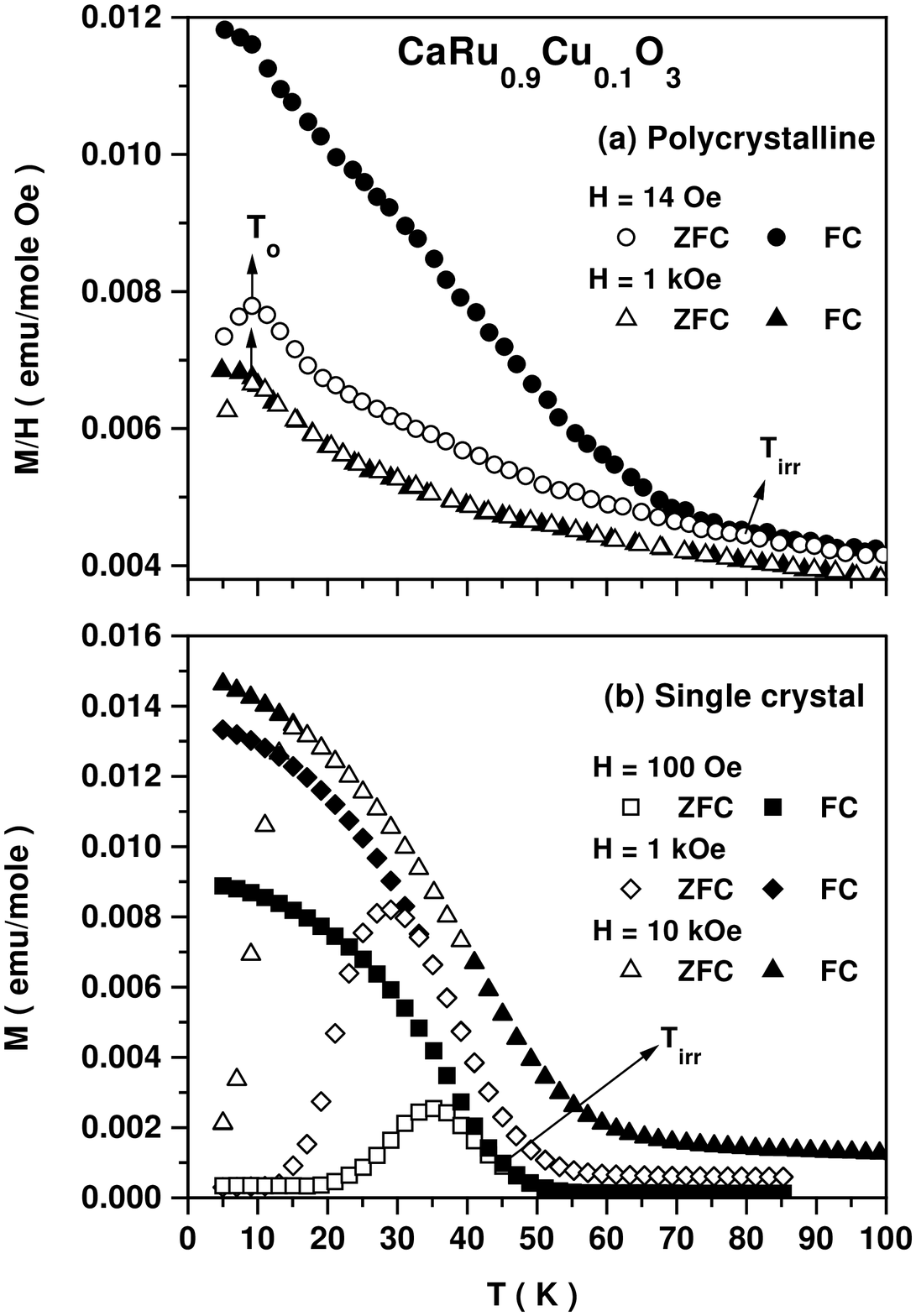,width=15cm}}

\caption
 {FC and ZFC curves for CaRu$_{0.9}$Cu$_{0.1}$O$_3$ at
different magnetic fields: (a) $\chi(T)$ vs $T$ for polycrystalline sample
and (b) $M(T)$ vs $T$ for single crystal ($H \bot c$-axis).}
\label{fig4}
\end{figure}

\begin{figure}[h]
\centerline{\psfig{file=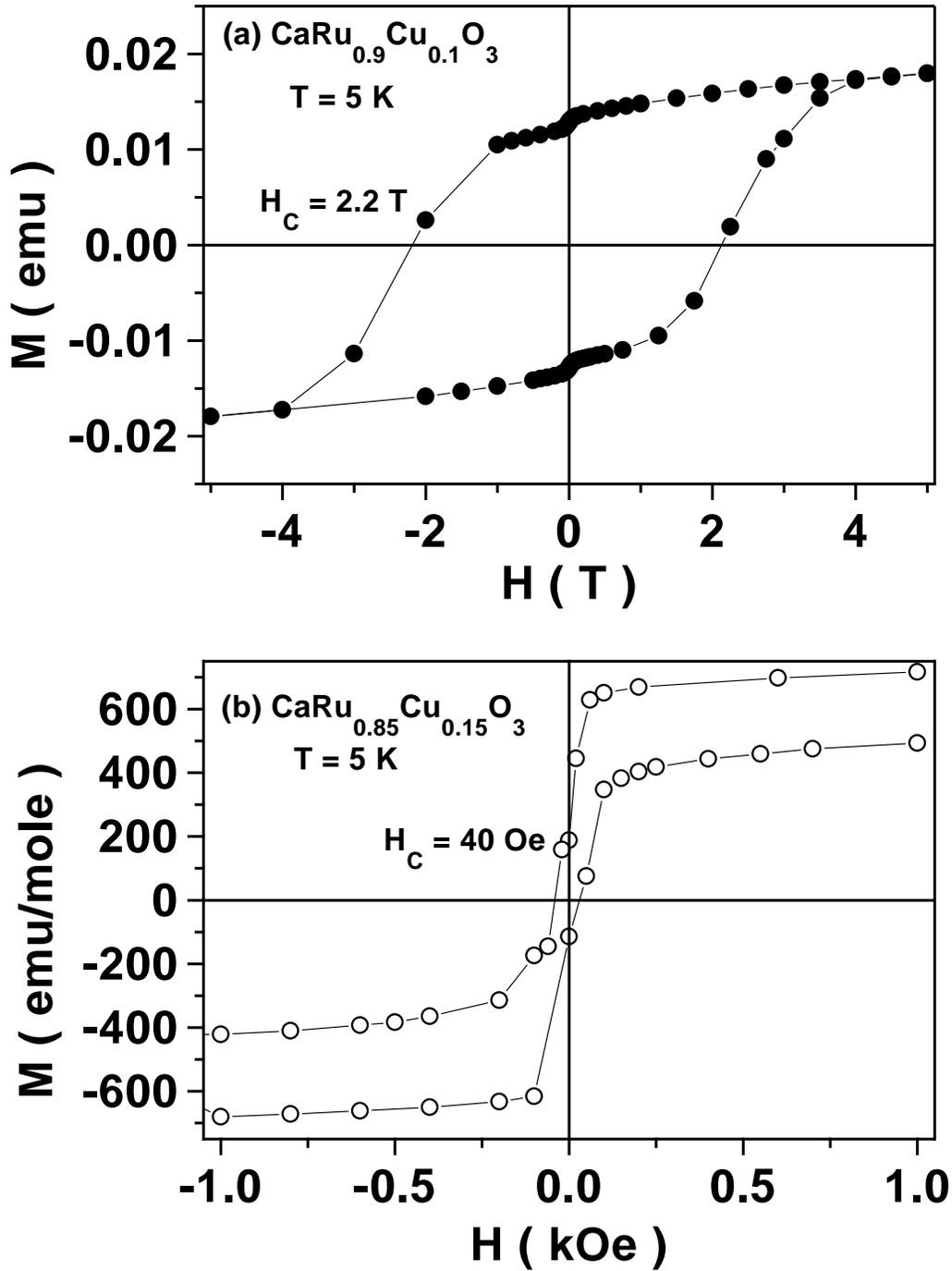,width=15cm}}

\caption
 {Magnetic hysteresis loops at $T=5\rm\,K$ (ZFC)
for single crystals: (a) CaRu$_{0.9}$Cu$_{0.1}$O$_3$
and  (b) CaRu$_{0.85}$Cu$_{0.15}$O$_3$. Field is applied perpendicular
to $c$-axis in both cases.}
\label{fig5}
\end{figure}

\begin{figure}[h]
\centerline{\psfig{file=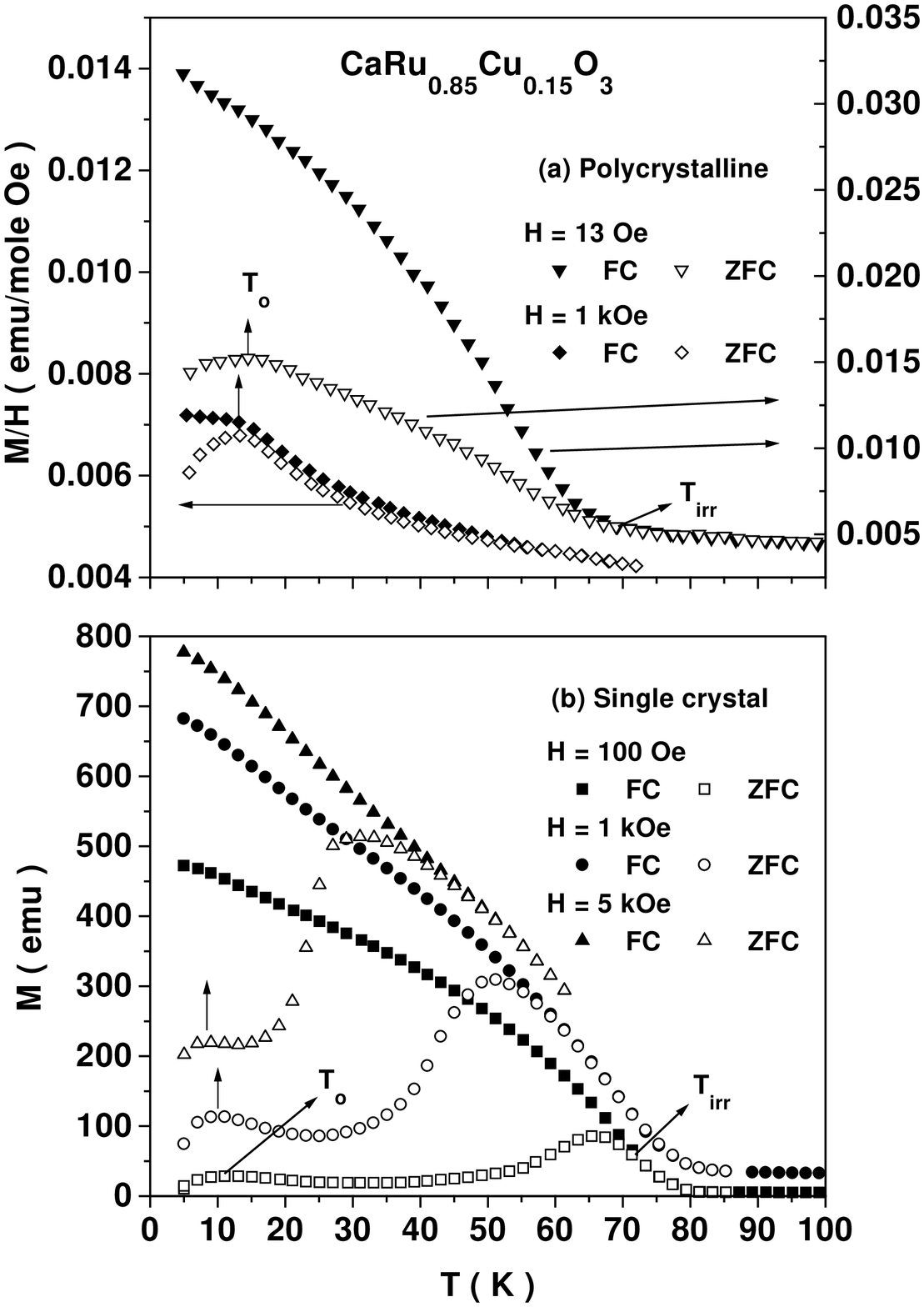,width=15cm}}

\caption
{ FC and ZFC curves for CaRu$_{0.85}$Cu$_{0.15}$O$_3$
 measured at different magnetic fields: (a) $\chi(T)$ vs $T$ for
 polycrystalline sample and (b) $M(T)$ vs $T$ for single crystal
($H \bot c$-axis).}
\label{fig6}
\end{figure}

\begin{figure}[h]
\centerline{\psfig{file=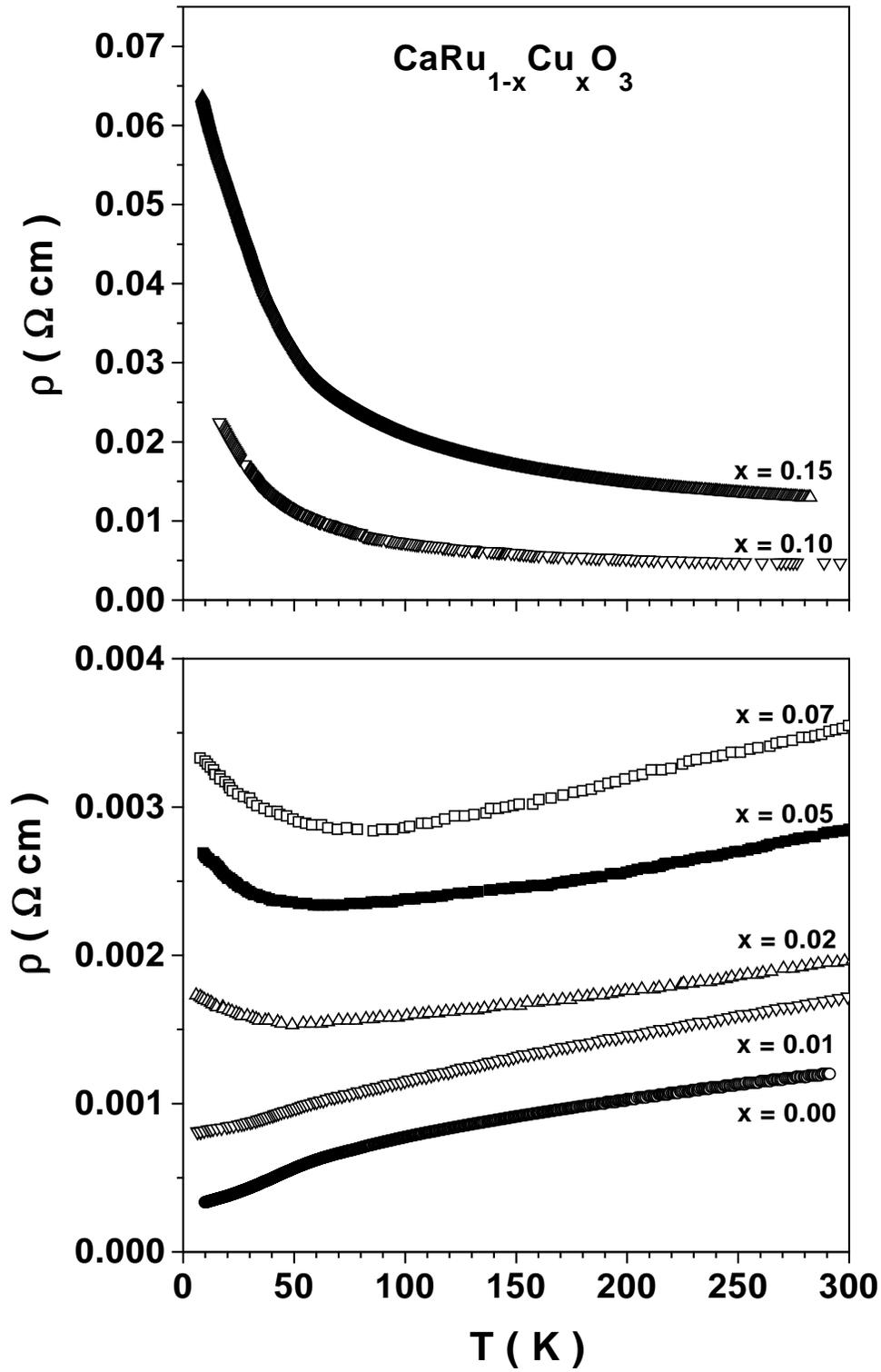,width=15cm}}

\caption{
 Electrical resistivity $\rho(T)$ vs temperature $T$
for polycrystalline CaRu$_{1-x}$Cu$_x$O$_3$ samples.}
\label{fig7}
\end{figure}

\vfill

\eject

\end{document}